\documentclass[preprint,superscriptaddress]{revtex4}
\usepackage{amsmath}
\usepackage{subfigure}

\usepackage{color}
\usepackage{graphicx}
\usepackage{epstopdf}
\makeatletter
\renewcommand\@biblabel[1]{#1.}
\makeatother
\newcommand{\eat}[1]{}

\begin{document}
\title{Predictability of road traffic and congestion in urban areas}

\author{Jingyuan Wang}\email{jywang@buaa.edu.cn}
\affiliation{School of Computer Science and Engineering, Beihang University,
Beijing, 100191, China}
\affiliation{Research Institute of Beihang University in ShenZhen, Sehnzhen, 518057 China}
\author{Yu Mao}
\affiliation{School of Computer Science and Engineering, Beihang University,
Beijing, 100191, China}
\author{Jing Li}
\affiliation{School of Computer Science and Engineering, Beihang University,
Beijing, 100191, China}
\author{Chao Li}
\affiliation{School of Computer Science and Engineering, Beihang University,
Beijing, 100191, China}
\affiliation{Research Institute of Beihang University in ShenZhen, Sehnzhen, 518057 China}
\author{Zhang Xiong}
\affiliation{School of Computer Science and Engineering, Beihang University,
Beijing, 100191, China}
\affiliation{Research Institute of Beihang University in ShenZhen, Sehnzhen, 518057 China}
\author{Wen-Xu Wang}\email{wenxuwang@bnu.edu.cn}
\affiliation{School of Systems Science, Beijing Normal University,
Beijing, 100875, China}

\begin{abstract}
Mitigating traffic congestion on urban roads, with paramount importance in urban development and reduction of energy consumption and air pollution, depends on our ability to foresee road usage and traffic conditions pertaining to the collective behavior of drivers, raising a significant question: to what degree is road traffic predictable in urban areas? Here we rely on the precise records of daily vehicle mobility based on GPS positioning device installed in taxis to uncover the potential daily predictability of urban traffic patterns. Using the mapping from the degree of congestion on roads into a time series of symbols and measuring its entropy, we find a relatively high daily predictability of traffic conditions despite the absence
of any a priori knowledge of drivers' origins and destinations and quite different travel patterns between weekdays and weekends. Moreover, we find a counterintuitive dependence of the predictability on travel speed: the road segment associated with intermediate average travel speed is most difficult to be predicted. We also explore the possibility of recovering the traffic condition of an inaccessible segment from its adjacent segments with respect to limited observability. The highly predictable traffic patterns in spite of the heterogeneity of drivers' behaviors and the variability of their origins and destinations enables development of accurate predictive models for eventually devising practical strategies to mitigate urban road congestion.

\end{abstract}

\maketitle

The past decades have witnessed a rapid development of modern society accompanied with an increasing demand for mobility in metropolises~\cite{mobility1,mobility2,mobility3,mobility4}, accounting for the conflict between the limits of road capacity and the increment of traffic demand reflected by severe traffic congestions~\cite{congestion1,congestion2}. Induced by such
problem, citizens suffer from the reduction of travel efficiency and the increase of both fuel consumption~\cite{consumption} and air pollution~\cite{pollution} related with vehicle emission. For instance, in recent years, a number of major cities in China have frequently experienced persistent haze, raising the need of better traffic management to mitigate congestion that is likely one of the main factors for the pollution~\cite{haze_China, uair}. Despite much effort dedicated to address the problems of traffic jam~\cite{Helbing_RMB}, urban planning~\cite{urban_planning1,urban_planning2} and traffic prediction~\cite{prediction1,prediction2,prediction3}, we still lack a comprehensive understanding of the dynamical behaviors of urban traffic. The difficulty stems from two factors: the lack of systematic and accurate data in conventional researches based on travel surveys and the diversity of drivers' complex self-adaptive behaviors in making routing choice decision~\cite{social_diversity}. Fortunately, ``big data" as the inevitable outcome in the information era opens new routes to reinvent urban traffic systems and offer solutions for increasingly serious traffic jams~\cite{big_data}. In this light, mobile phone data have been employed to explore road usage patterns in urban areas~\cite{phone_data1,phone_data2,phone_data3}. However, to eventually implementing control on road traffic, predicting traffic conditions is the prerequisite, which prompts us to wonder, to what degree traffic flow on complex road networks is predictable with respect to high self-adaptivity of drivers and without any a priori knowledge of their origins and destinations.

In this paper, we for the first time explore the predictability of urban traffic and congestion by using comprehensive records of Global Position System (GPS) devices installed in vehicles. The data provide the velocity and locations of a large number of taxis in real time, enabling investigation and quantification of the predictability of segments in main roads in an urban road network. In particular, we establish a mapping from the degree of congestion on a segment of road into a time series of symbols, which allows us to exploit tools in the information theory, such as entropy~\cite{Entropy} and Fano's inequality~\cite{Fano} to measure the predictability of traffic condition on a segment of road. Our methodology is inspired by the seminal work of Song et al. who incorporate information theory into time series analysis to measure the limited predictability of individual mobility~\cite{predictability}. Our main contribution is that we extend the tools of time series analysis to the collective dynamics of road traffic rather than at individual level, by mapping the vehicle records from GPS into road usage so as to offer the predictability of traffic conditions at different locations. In contrast to the traditional way based on origin-destination analysis~\cite{OD_analysis}, our approach relies only on short-time historical records of traffic conditions without the need of a priori knowledge of drivers' origins and destinations and their associated navigation strategies. Our accessibility of such individual-level information is inherently limited by the diversity in population, job switching, moving and urbanization. Our research yields a number of interesting findings, including relatively high daily predictability of traffic conditions in the three Ring Roads in Beijing~\cite{rings} despite quite different travel patterns on the weekends compared to working days, the non-monotonic dependence of the predictability on vehicle velocity and the recoverability of the traffic condition of an inaccessible segment by the information of its adjacent observable segments. Thus we present a general and practical approach for understanding the predictability of real time urban road traffic and for devising effective control strategies to improve the roads' level of service.

\noindent
\section{Results}

\begin{figure}[t]
\includegraphics[width=0.9\columnwidth]{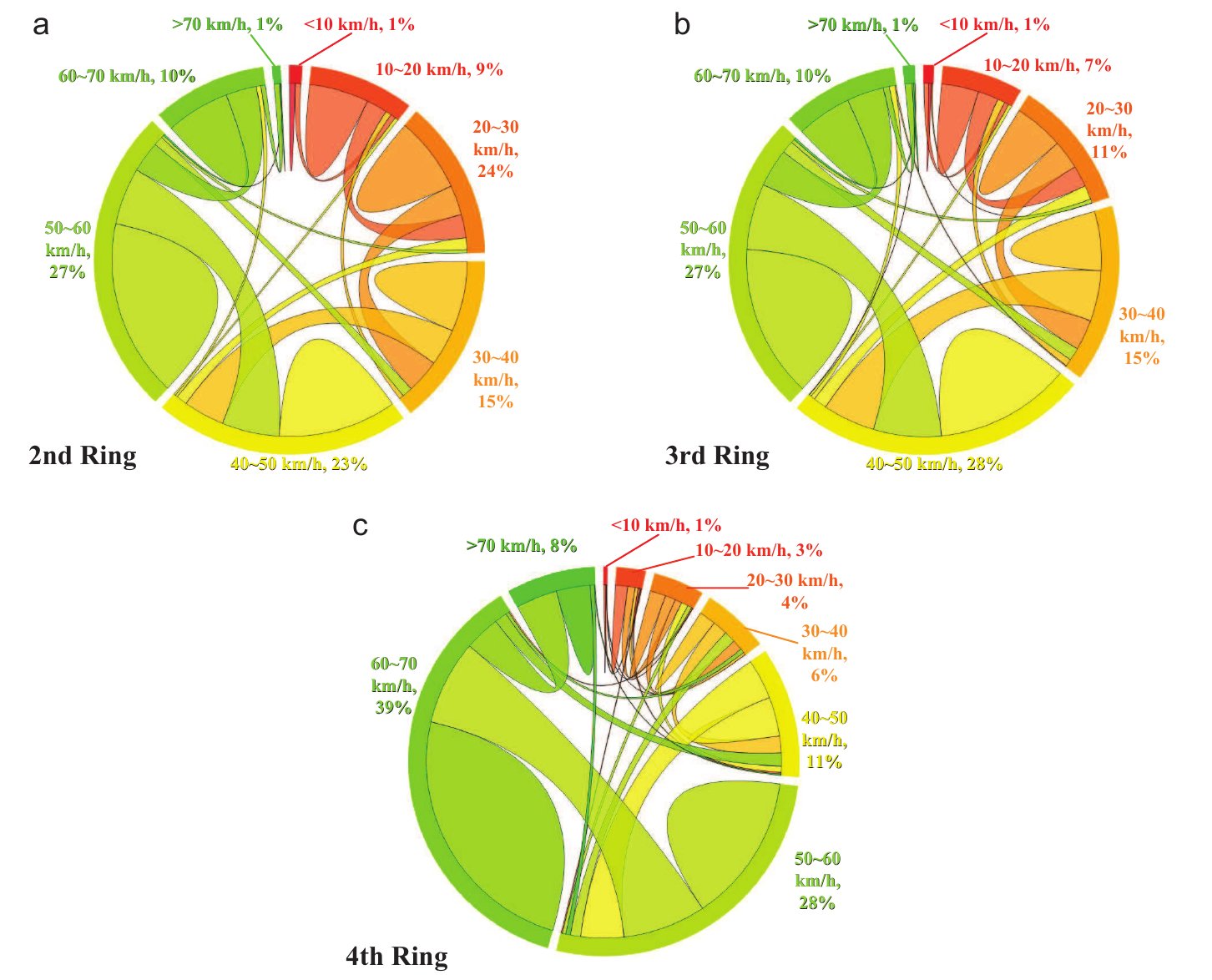}\\
\caption{{\bf Transition probability of speed states.} (a)-(c) Transition probability between different speed states in the 2nd (a), 3rd (b) and 4th (c) Ring Roads of Beijing. The speed $V$ between 10km/h and 70km/h is divided into 6 states with equal speed interval $\Delta V=10$km/h. Due to rare observations for $V<10$km/h and $V>70$km/h, they are set to be two states respectively, without any further partitions. For each Ring Road, the result is obtained by averaging over all road segments with equal length $\Delta L=1$km. We see that for each state, remaining unchanged and shifting to its adjacent states constitute a very large proportion, implying a potential stable regulation in the traffic patterns.
}\label{Fig:1}
\end{figure}

We explore the predictability of traffic conditions by using the GPS records of more than 20000 taxis in Beijing, China, (see Methods for data description and processing). We focus on the three Ring Roads, the 2nd, 3rd and 4th Rings in Beijing by mapping the states of vehicles into the traffic conditions on the roads. The three rings bear the most heavy traffic burden in Beijing and the data records pertains to them with high frequency are sufficient for quantifying their traffic conditions. In particular, we divide each ring into a number of segments with given $segment~length~\Delta L$, and measure the traffic condition of each segment by the average velocity of vehicles. To simplify our study, we discretize the average velocity of the segments in the range from 0km/h to the speed limit 100km/h with a certain $speed~level~interval~\Delta V$, e.g., 10km/h. Thus, the mapping gives rise to a time series of discrete states of speed for each road segment, which allows us to do some analysis of discrete time series to reveal intrinsic traffic patterns. The dynamical behavior of a whole ring can then be quantified by that of all segments of it.
Figure~\ref{Fig:1} shows the transition probabilities between different ranges of speed, say, speed states. We find that on average, a speed state is more likely to remain unchanged or shift to its nearby states rather than change to a distant state. These observations imply the existence of a potentially stable transition pattern that may facilitate the prediction of traffic conditions and congestion from historical records. \eat{We also found that the transition pattern of the clockwise direction is approximately the same as that of the anticlockwise direction for all the three ring roads. Therefore, without loss of generality, in the study below, we only consider the clockwise direction.}

We exploit information entropy~\cite{Entropy} to quantify the uncertainty of speed transition and the degree of predictability characterizing the time series of the speed at each segment. By following Ref.~\cite{predictability}, we assign three entropy measures to each road segment's traffic pattern: (i)~{\em Random Entropy} $S_{i}^\text{rand}$. Random entropy is defined as $S_{i}^\text{rand}=\log_{2}N_{i}$ where $N_{i}$ is the number of distinct states, or speed levels, reached by road segment $i$. (ii)~{\em Temporal-uncorrelated Entropy} $S_{i}^\text{unc}$. Temporal-uncorrelated entropy is defined as $S_{i}^\text{unc}=-\sum_{j=1}^{N_{i}}p_{i}(j)\log_{2}p_{i}(j)$, where $p_{i}(j)$ is the probability that the state $j$ is reached by the road segment $i$. (iii)~{\em Actual Entropy $S_{i}$}. Actual entropy is defined as $-\sum_{T_{i}^{'}\subset T_{i}}P(T_{i}^{'})\log_{2}[P(T_{i}^{'})]$, where $T_{i}=\{X_{1}, X_{2}, ..., X_{L}\}$ denotes the sequence of states that road segment $i$ reaches in observation. $P(T_{i}^{'})$ is the probability of finding the time-ordered subsequence $T_{i}$ in the state transition sequence of segment $i$. It is noteworthy that the random entropy $S^\text{rand}_{i}$ reflects the degree of predictability of a road segment's state transition based on the assumption that each state is visited with equal probability. For the temporal-uncorrelated entropy $S_{i}^\text{unc}$, it takes the heterogeneity in the probability into account, but omits the order of the transition. In contrast, the actual entropy $S_{i}$ by considering both heterogeneous probability and temporal correlation offers more realistic characterization of the traffic patterns.

\begin{figure}[t]
\includegraphics[width=0.9\columnwidth]{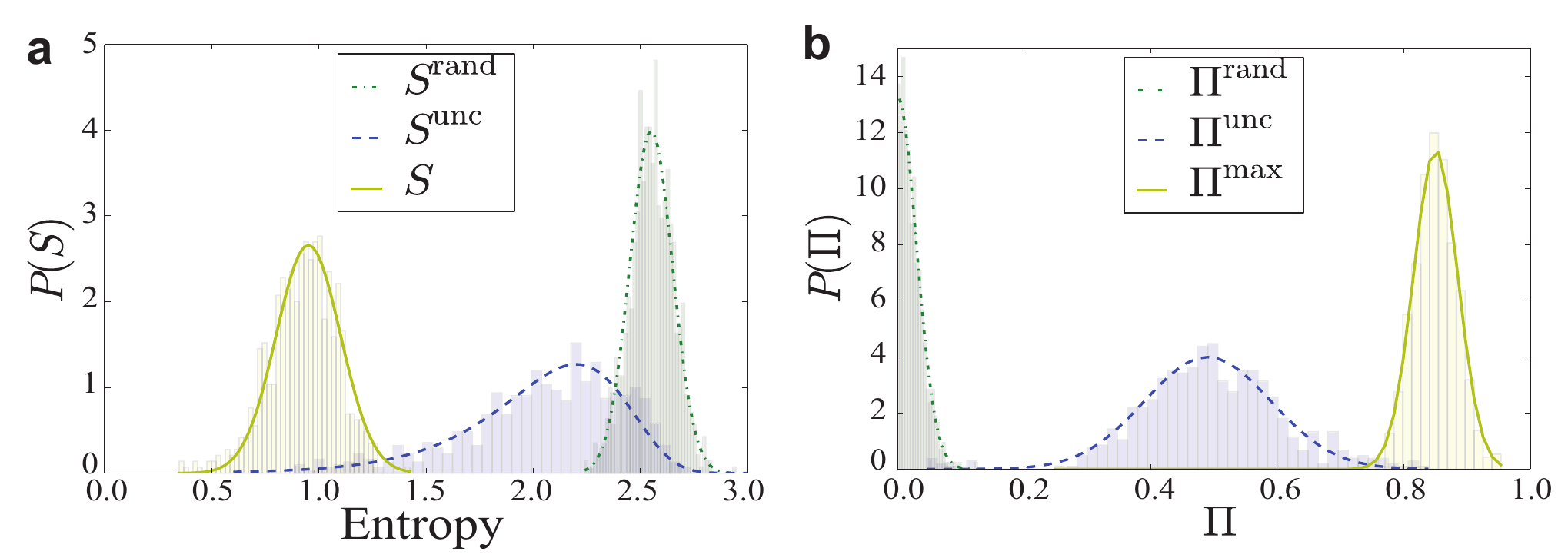}\\
\caption{{\bf Distributions of entropy and probability $\Pi$.} (a) The distribution of the random entropy $S^\text{rand}$, the uncorrelated entropy $S^\text{unc}$
and the entropy $S_i$ of road segments in the 2nd Ring Road in Beijing. (b) The distribution of the $\Pi^\text{rand}$, the $\Pi^\text{unc}$ and
the $\Pi^{\max}$ across all road segments. The road segments are of identical length $\Delta L = 1$km and the interval of speed state is $\Delta V = 10$km/h.
The 3rd and 2th Ring Roads show similar results of $P(S)$ and $P(\Pi)$ to that of the 2nd Ring Roads.
}
\label{Fig:2}
\end{figure}

The sufficient data with high record frequency on the three ring roads allow us to calculate the actual entropy $S_i$ that in principle requires a continuous record of a road segment's momentary state. As shown in Fig.~\ref{Fig:2}(a), we can see remarkable difference between $P(S)$ and $P(S^\text{rand})$. To be concrete, $S^\text{rand}$ peaks at about $2.6$, indicating that on average each update of the speed state represents 2.6 bits per hour new information. In other words, the new speed level could be found in average $2^{2.6}\approx 6$ states. In contrast, the fact that $P(S)$ of the actual entropy peaks at $S=0.9$ demonstrates that the real uncertainty in a segment's speed state is $2^{0.9}\approx 1.87$ rather than 6.

The entropy of a segment's speed allows us to measure the predictability $\Pi$ that a suitable predictive algorithm can correctly predict the segment's future speed state. In analogy with Ref.\cite{predictability}, the predictability measure is subject to Fano's inequality. Specifically, if the speed level of a single road segment is updated in $N$ states with the time, then its predicability $\Pi\le\Pi^{\max}(S,N)$, where $\Pi^{\max}$ could be acquired by solving $$S=H(\Pi^{\max})+(1-\Pi^{\max})\log_{2}(N-1),$$ where $H(\Pi^{\max})$ represents the binary entropy function, namely $$H(\Pi^{\max})=-\Pi^{\max}\log_{2}(\Pi^{\max})-(1-\Pi^{\max})\log_{2}(1-\Pi^{\max}).$$ For a road segment with $\Pi^{\max}=0.1$, we could predict its state transition accurately only in $10\%$ of the cases. An equivalent statement is that $10\%$ is the upper bound of probability for any algorithms attempting to predict the segment's speed state transition. Since we calculates $\Pi^{\max}$ base on $S^\text{rand}$, $S^\text{unc}$ and $S$, the result is encouraging. We found that under the condition $\Delta L = 1km$ and $\Delta V=20km/h$, the predicability of the 2nd Ring Road segments is narrowly peaked approximately at 0.83, indicating that it is theoretically possible to predict the transition of speed status in $83\%$ of the cases. This high predictability with bounded distribution indicate that, despite the diversity of drivers' origins, destinations, their routing decisions and adaptive behaviors, strikingly the traffic patterns as a collective behavior of a large number of drivers are of high degree of potential predictability exclusively based the historical records of daily traffic patterns in the absence of any individual level information. We have also explored the maximum predictability $\Pi^\text{unc}$ and $\Pi^\text{rand}$ based on $S^\text{unc}$ and $S^\text{rand}$, as shown in Fig.~\ref{Fig:2}(b). We see that both maxima in $P(\Pi^\text{unc})$ and $P(\Pi^\text{rand})$ are much lower than that of $P(\Pi^\text{max})$, manifesting that $\Pi^\text{max}$ is a much better predictive tool than the other two and the temporal order of traffic patterns contains significant information for precisely predicting future patterns.

\begin{figure}[t]
\includegraphics[width=0.9\columnwidth]{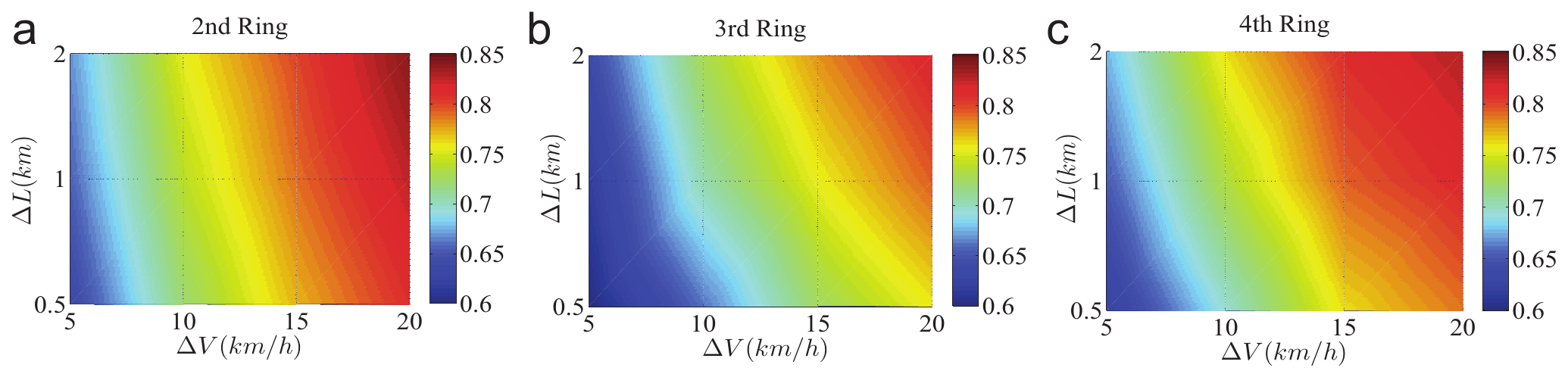}\\
\caption{{\bf Predictability of the three Ring Roads.} (a)-(c) The dependence of the maximum value $\Pi^{\max}$ on $\Delta L$ and $\Delta V$ for the 2nd (a), the 3rd (b) and the 4th (c) Ring Road. The color bars represent
the values of $\Pi^{\max}$. The results for each Ring Road are the average over all road segments in the Ring Road.
}\label{Fig:3}
\end{figure}

We further explore how the settings of the road segment length $\Delta L$ and speed level interval $\Delta V$ affect the predictability. As shown in Fig.~\ref{Fig:3}, except very small $\Delta V$ and very short $\Delta L$, quite high average predictability is observed. This provides strong evidence for the generally high predictability of traffic conditions of the three ring roads. The relatively low predictability for extreme cases is ascribed to the relatively big fluctuations in the average speed resulting from insufficient records. For example, for a road segment with very short length, the probability of finding a taxi in it within a certain time interval will be low. In other words, in this scenario, the data records of taxis will become insufficient to capture the actual average speed in the segment, accounting for the big fluctuation of speed and inaccurate reflection of the traffic pattern in the segment. Similarly, for small $\Delta V$, the insufficient data subject to each speed state is incapable of characterizing the real situation, leading to the specious low predictability. Nevertheless, based on our findings, insofar as the records are adequate to measure traffic conditions, the traffic patterns are highly predictable, regardless of the settings of the road segment length and speed interval.

\begin{figure}
\includegraphics[width=0.9\columnwidth]{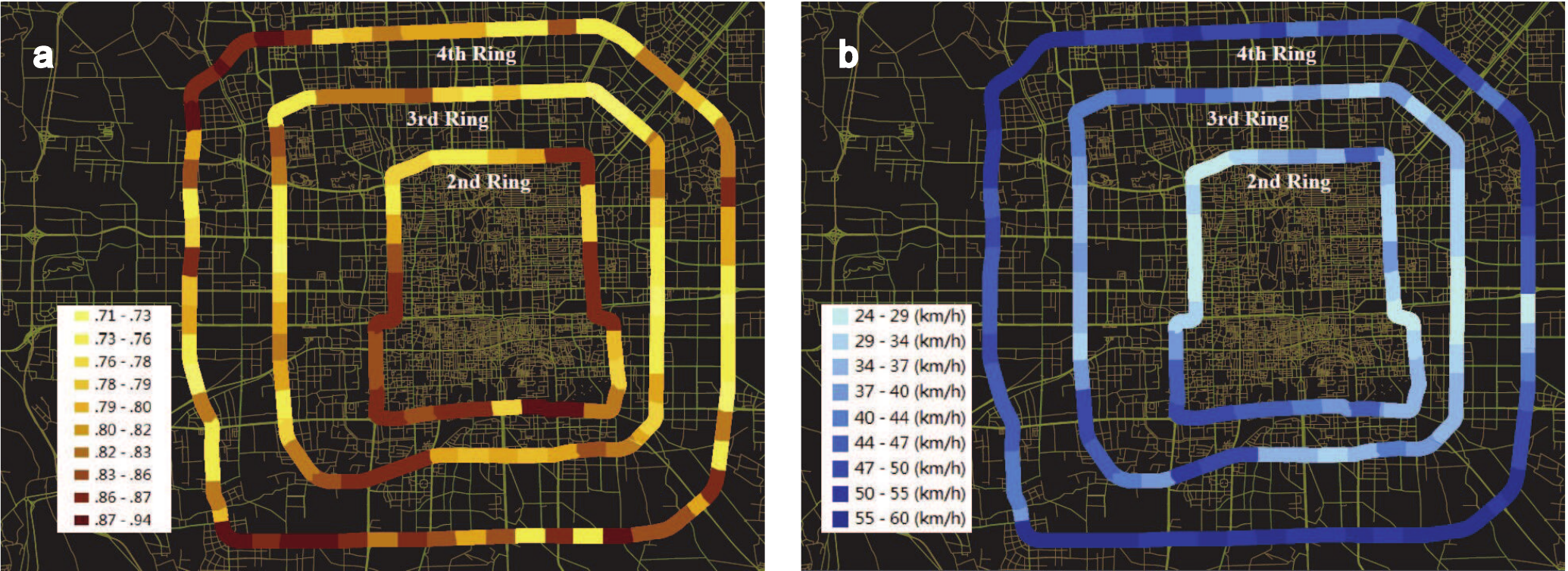}\\
\caption{{\bf Local predictability and average speed.} (a) The local predictability of road segments in the three Ring Roads.
(b) The local average speed of road segments in the three Ring Roads. In (a), the color bar represents the maximum value
$\Pi^{\max}$ of road segments and In (b), the color bar represents the average speed of road segments.
}
\label{Fig:4}
\end{figure}

\begin{figure}
\includegraphics[width=\columnwidth]{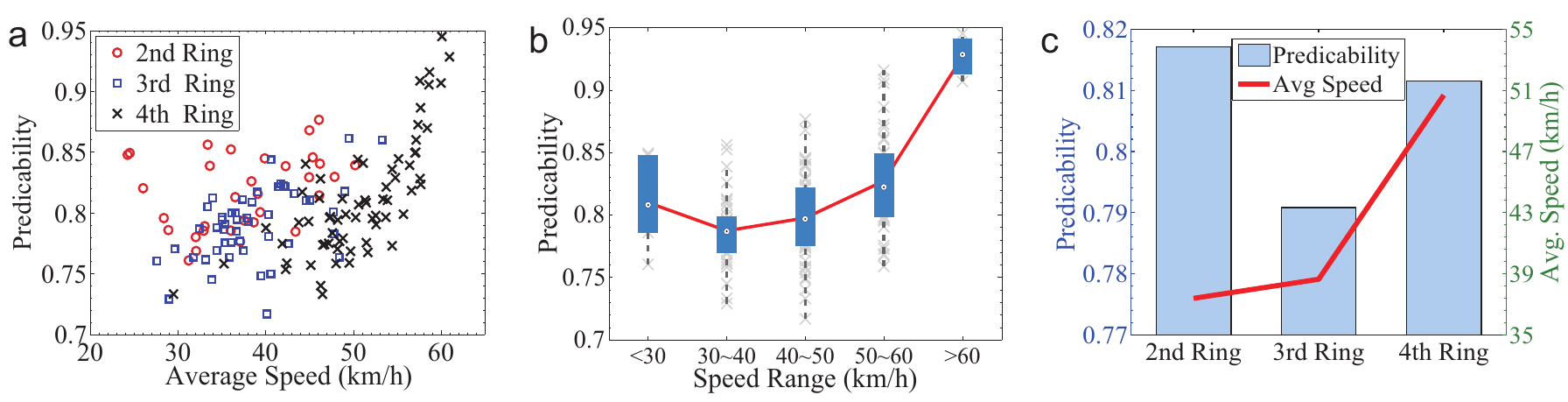}\\
\caption{{\bf Relationship between predictability and average speed.} (a) Predictability as measured by $\Pi^{\max}$ as a function of the average speed
for the three Ring Roads. (b) Box plots of the predictability in different ranges of the average speed. (c) The predictability and the average
speed of each entire Ring Road. The results are obtained for $\Delta L =1$Km and $\Delta V=10$km/h. the highest and lowest values outside the
boxes represent $9\%$ and $91\%$ of the rank of predicted values, respectively, the upper and lower bound of
the boxes represent $25\%$ and $75\%$ of the rank of predicted values, respectively, and the bars inside the
boxes characterize the median value.
}\label{Fig:5}
\end{figure}

Although the traffic patterns of the three ring roads on average are highly predictable, there are certain variations between different segments. Figure~\ref{Fig:4}(a) shows the local predictability of each segment on the map. We find that the local predictability is correlated with the average local speed (Fig.~\ref{Fig:4}(b)), prompting us to investigate the correlation between them. Interestingly, we observe a non-monotonic correlation between the local predictability and average speed with the lowest predictability arising at intermediate speed, as shown in Fig.~\ref{Fig:5}(a) and~\ref{Fig:5}(b). As a result, we also find that it is most difficult to predict the traffic conditions of the 3rd ring road, due to its intermediate average speed compared to the 2nd and 4th ring roads. A heuristic explanation for this phenomenon can be provided with respect to the variational direction of speed. Suppose that in a segment all the vehicles are fully stopped because of heavy congestion. One minute later, remaining stopped or starting to pull away are the only two possible situations. Let's consider another extreme case in which all vehicles are moving along the speed limit of a road without any congestion. One minute later, there are also only two possible scenarios, i.e., their speeds remain unchanged or reduce because of some suddenly emerged congestion. In contrast to the extreme cases, for a car with intermediate speed, it may accelerate, decelerate or keep its current speed some time later, relying on what happens in the near future. Therefore, due to higher variant possibilities of intermediate speed compared to that of low and high speed, the traffic condition of a segment with intermediate average speed is relatively most difficult to be predicted.

\begin{figure}
\includegraphics[width=0.8\columnwidth]{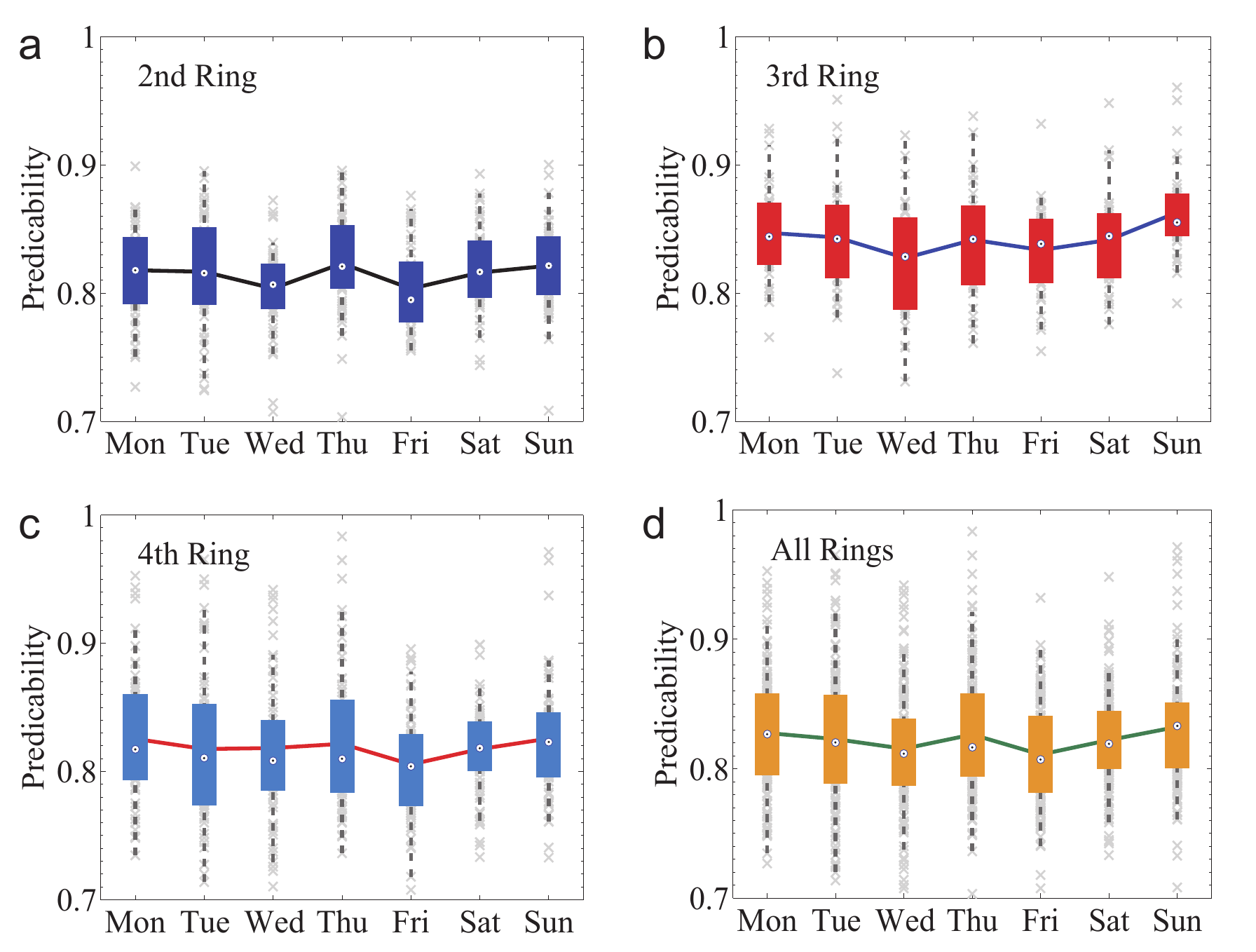}\\
\caption{{\bf Daily predictability.} (a)-(c), The daily predictability during a week of the 2nd (a), 3rd (b) and 4th (c) Ring Road.
(d) The daily predictability averaging over all of the three Ring Roads during a week. The parameter values and the box plots are the same as in Fig.~\ref{Fig:5}.
}\label{Fig:6}
\end{figure}

To gain a deeper understanding of the predictability of traffic patterns, we explore the effect of commuter demand on daily traffic predictability in terms of the comparison between weekdays and weekends. It is intuitive that the commuter demand during weekdays may induce quite different traffic patterns and congestion distributions compared to that on weekends. However, to our surprise, despite the obvious difference, we find that the daily traffic patterns in a week are of very similar predictability, nearly regardless of the commuter demand, as shown in Fig.~\ref{Fig:6}. These striking results suggest that both weekdays and weekends have their specific inherent patterns encoded in the historical records, accounting for the relatively high and similar predictability.

Next, we explore the probability of inferring the state of a segment from the state series of its adjacent segments. This problem is related to the observability that in the control theory is defined as if a system's state can be fully recovered from a set of observable quantities~\cite{observability}. To the urban road traffic, inferring traffic conditions at some locations from the observation of the other segments has important applications in monitoring and controlling traffic in real time from a limited number of speed detectors. In analogy with the predictability, we calculate the inference probability $\widetilde{\Pi}$ of a segment based on the information entropy and the Fano's inequality. However, different from the predictability, here the information entropy is calculated by $S^{'}_{i}=-\sum_{R_{i}^{'}\subset R_{i}}P(R_{i}^{'})\log_{2}[P(R_{i}^{'})]$, where $R_{i}=\{X_{1}, X_{2}, ..., X_{L}\}$ denotes the states observed within a single time interval of $L$ road segments connected in a sequence, and $P(R_{i}^{'})$ is the probability of finding the subsequence $R_{i}^{'}$ in this sequence. Similarly, by solving $S^{'}=H(\widetilde{\Pi}^{\textnormal{max}})+(1-\widetilde{\Pi}^{\textnormal{max}})\log_{2}(N-1)$, we obtain an upper bound $\widetilde{\Pi}^{\textnormal{max}}$ which captures the inference probability of the traffic pattern of a road segment from its observable adjacent segments.

\begin{figure}
\includegraphics[width=0.5\columnwidth]{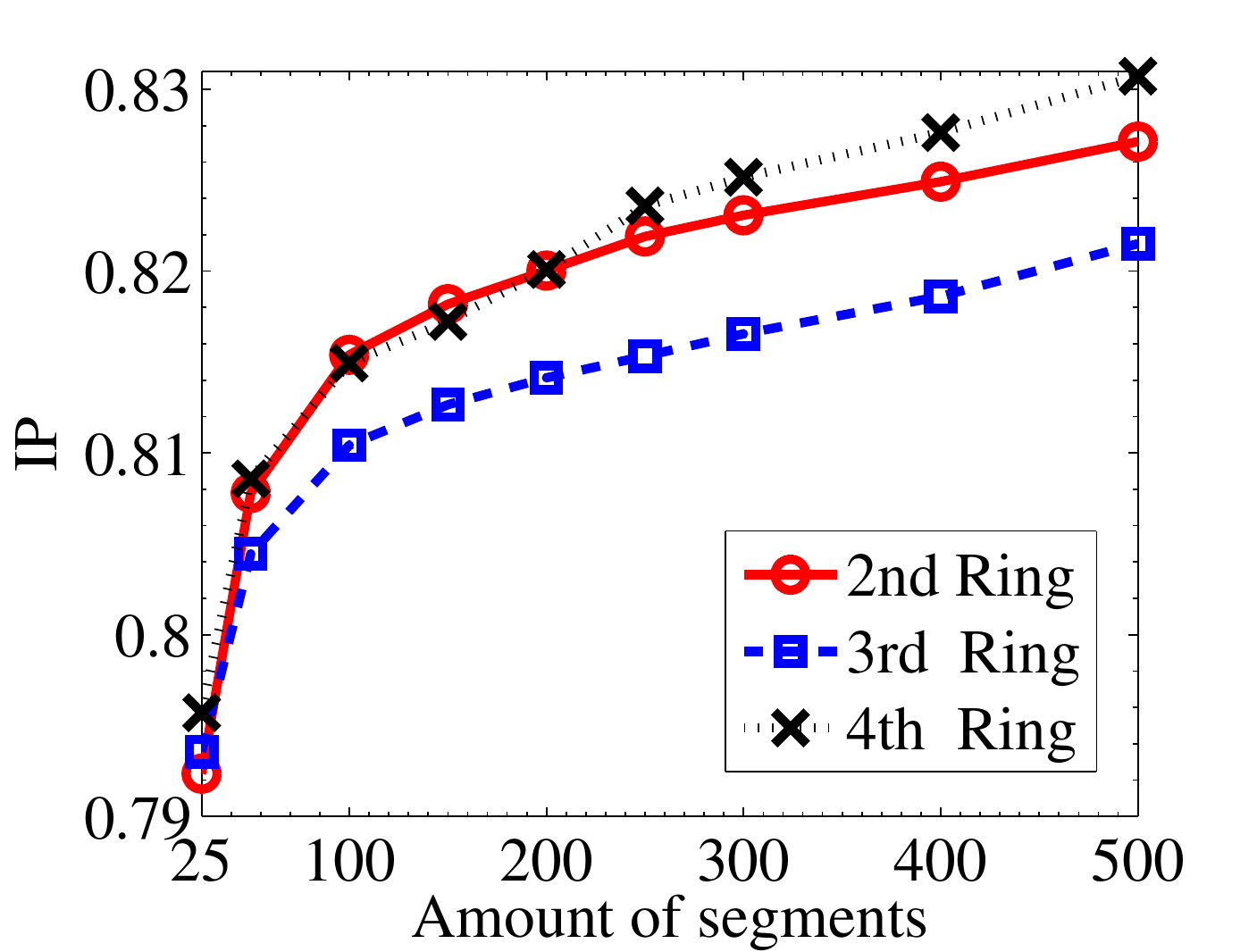}
\centering
\caption{{\bf Inference probability.} The inference probability (IP) as a function of the amounts of segments for
all the three Ring Roads. Here IP is measured by the upper bound of $\widetilde{\Pi}^{\max}$.}
\label{Fig:7}
\end{figure}

As shown in Fig.~\ref{Fig:7}, we see that the inference probability increases as the amount of segments increases for all the three ring roads.
This phenomenon can be heuristically explained as follows. For sufficiently short segment lengths (sufficient number of segments), the average vehicle speed in a segment will be sufficiently close to that in its adjacent segments, enabling an accurate inference of the segment's state by trivially using that in its neighborhood. The increment of segment length induces more difference between adjacent segments, rendering the inference more difficult. As a result, the inference probability is an increase function of the amount of segments. More importantly, our results provide a quantitative understanding of the inference probability in terms of the number of segments, which is valuable for determining the density of speed detectors installed so as to infer the traffic conditions of the entire road in real time with a certain accuracy.
In addition, we also find that the inference probability of the 3rd ring road exhibits the lowest values compared to the 2nd and 4th ring roads, which is the same as the predictability rank of the three ring roads, i.e., the 3rd ring road is of the lowest predictability. This suggests that the average vehicle speed plays similar roles in both predictability and inference probability, which deserves deeper explorations.

\noindent
\section{Discussions}

In summary, using the GPS records of vehicles to capture the traffic patterns on urban roads in the combination of entropy and Fano's inequality demonstrates that daily traffic patterns in the three major ring roads in Beijing are highly predictable by relying only on short-time historical records, without any a priori knowledge of drivers' origins and destinations, driving habits, navigation strategies, and adaptive behaviors. We have also found that despite the apparently different traffic patterns on weekdays from that on weekends, where the former is highly affected by commuter demand, their traffic patterns exhibit similarly high predictability. This result indicates that each day has its specific inherent regularity and traffic pattern encoded in the historical records. Another striking finding is that the local predictability is non-monotonically correlated with the average velocity and the lowest predictability arises at intermediate velocity. Consequently, the traffic conditions of the 3rd ring road due to its intermediate average velocity compared to the 2nd and 4rd ring roads, is most difficult to be predicted. We have provided a heuristic explanation for this counterintuitive phenomenon. Furthermore, the probability of inferring the traffic pattern of an inaccessible road segment from the state series of its adjacent segment is explored by using entropy and Fano's inequality, which is important for monitoring the traffic condition of the entire road network with respect to the limits of our ability to observe every location in real time.

All of these findings are valuable for the development of predictive models and algorithms for achieving actual predictions of traffic conditions in real time based solely on short-time historical records, without the need of individual-level information that in principle is impossible to be fully accessed. Relying on the successful prediction of traffic patterns, it is feasible to implement effective control to release and prohibit congestion by exploiting traditional approaches in traffic engineering~\cite{traffic_engineering} and the recently developed controllability theory for complex networks~\cite{Control1,Control2}. Urban road network as a typical complex networked system exhibits a variety of dynamical behaviors, such as the phantom jam and the diffusion of congestion~\cite{Helbing_RMB}. Thus, it is imperative to control the road network as a whole in virtue of the controllability framework rather than controlling a single road or area individually. Our approach gains new insight into mitigating increasingly severe congestion in urban areas by combining ``big data" and the tools in information theory and for time series analysis. Further effort, we hope, will be inspired toward predicting traffic patterns and devising effective strategies to alleviate traffic congestion in urban areas.

\section{Methods}
We use OpenStreetMap~\cite{OSM} to extract all roads in the spatial range of Beijing from available database. We then retrieve the trajectories of vehicles. The data set that we used contains the trajectories of 20000 taxies recorded every minutes within a month in Beijing. For each record, the location (the latitude and longitude), the direction, the state (whether there are any passengers in the taxi), the time stamp and the velocity updated in every minutes are included. Because of the inevitable error in the GPS locating process, all the records are preprocessed to match the GPS trajectories to the road by exploiting the ST-Matching algorithm~\cite{gps_mapping}. After that, each GPS record is mapped to a road segments of OpenStreetMap.

To be concrete, ST-Matching algorithm of Ref.~\cite{gps_mapping} is implemented via four steps: (i) {\it Candidate Preparation.} Firstly, for each GPS record point, the ST-Matching algorithm retrieves a set of candidate road segments within a fixed radius $r$, which is set to be 20 meters. For the points without any candidates within $r$, the algorithm discards them as invalid records. (ii){\it Spatial Analysis.} The algorithm next evaluates the given candidate segments by using ``observation probability'' and ``transmission probability'' to express the geometric and topological information of each candidate segments and the spatial relationship between them. This step gives rise to the spatial analysis function $F_{s}(c_{i-1}^{t}\to c_{i}^{s})$, which is simply the product of the observation probability and transmission probability. In this function, $c_{i}^{s}$ represents the $s$th candidate segment of the $i$th GPS sampling record. This function measures the probability that the the $i$th record is on $c_{i}^{s}$, given an assumed real segment mapping of the $(i-1)$th record, that is $c_{i-1}^{t}$. (iii) {\it Temporal Analysis.} The ST-Matching algorithm exploits the temporal analysis function $F_{t}(c_{i-1}^{t}\to c_{i}^{s})$ to further incorporate the temporal features into the map-matching process. This step is available for the situation that only spatial analysis could not handle. Specifically, if the trajectory of a vehicle lies between a freeway and a service road, and it moves in a relatively high speed, then more likely it is that the vehicle is on the freeway. (iv) {\it Result Matching.} Finally, after $F_{s}(c_{i-1}^{t}\to c_{i}^{s})$ and $F_{t}(c_{i-1}^{t}\to c_{i}^{s})$ is computed, the algorithm uses the ST-function to evaluate each candidate segments, that is $F(c_{i-1}^{t}\to c_{i}^{s})$=$F_{s}(c_{i-1}^{t}\to c_{i}^{s})\times F_{t}(c_{i-1}^{t}\to c_{i}^{s}), 2\leq i\leq n$. Thus, the problem is converted to finding a path with the highest ST-function value, given the candidates for all sampling points.

After the map-matching process, each point is assigned with an attribute which represents the road segment that the point is on. Based on the work before, we could generate the time series of each road segment's speed states.



{

}

\end{document}